\documentclass[twocolumn,11pt]{article}
\usepackage{amssymb,amsmath,latexsym}

\usepackage[square,numbers]{natbib}
\usepackage{chemarr}
\usepackage{enumitem}
\usepackage{graphicx}
\usepackage{amssymb}
\usepackage{lineno}
\usepackage{amsmath}
\usepackage{sidecap}
\usepackage{wrapfig}
\usepackage{multirow}
\usepackage{float}
\begin{document}
\date{\today}
\newcommand{\MSun}{{M_\odot}}
\newcommand{\LSun}{{L_\odot}}
\newcommand{\Rstar}{{R_\star}}
\newcommand{\calE}{{\cal{E}}}
\newcommand{\calM}{{\cal{M}}}
\newcommand{\calV}{{\cal{V}}}
\newcommand{\calO}{{\cal{O}}}
\newcommand{\calH}{{\cal{H}}}
\newcommand{\calD}{{\cal{D}}}
\newcommand{\calB}{{\cal{B}}}
\newcommand{\calK}{{\cal{K}}}
\newcommand{\labeln}[1]{\label{#1}}
\newcommand{\Lsolar}{L$_{\odot}$}
\newcommand{\farcmin}{\hbox{$.\mkern-4mu^\prime$}}
\newcommand{\farcsec}{\hbox{$.\!\!^{\prime\prime}$}}
\newcommand{\kms}{\rm km\,s^{-1}}
\newcommand{\cc}{\rm cm^{-3}}
\newcommand{\Alfven}{$\rm Alfv\acute{e}n$}
\newcommand{\Vap}{V^\mathrm{P}_\mathrm{A}}
\newcommand{\Vat}{V^\mathrm{T}_\mathrm{A}}
\newcommand{\D}{\partial}
\newcommand{\DD}{\frac}
\newcommand{\TAW}{\tiny{\rm TAW}}
\newcommand{\mm }{\mathrm}
\newcommand{\Bp }{B_\mathrm{p}}
\newcommand{\Bpr }{B_\mathrm{r}}
\newcommand{\Bpz }{B_\mathrm{\theta}}
\newcommand{\Bt }{B_\mathrm{T}}
\newcommand{\Vp }{V_\mathrm{p}}
\newcommand{\Vpr }{V_\mathrm{r}}
\newcommand{\Vpz }{V_\mathrm{\theta}}
\newcommand{\Vt }{V_\mathrm{\varphi}}
\newcommand{\Ti }{T_\mathrm{i}}
\newcommand{\Te }{T_\mathrm{e}}
\newcommand{\rtr }{r_\mathrm{tr}}
\newcommand{\rbl }{r_\mathrm{BL}}
\newcommand{\rtrun }{r_\mathrm{trun}}
\newcommand{\thet }{\theta}
\newcommand{\thetd }{\theta_\mathrm{d}}
\newcommand{\thd }{\theta_d}
\newcommand{\thw }{\theta_W}
\newcommand{\beq}{\begin{equation}}
\newcommand{\eeq}{\end{equation}}
\newcommand{\bsplit}{\begin{split}}
\newcommand{\esplit}{\end{split}}
\newcommand{\ben}{\begin{enumerate}}
\newcommand{\een}{\end{enumerate}}
\newcommand{\bit}{\begin{itemize}}
\newcommand{\eit}{\end{itemize}}
\newcommand{\barr}{\begin{array}}
\newcommand{\earr}{\end{array}}
\newcommand{\bc}{\begin{center}}
\newcommand{\ec}{\end{center}}
\newcommand{\DroII}{\overline{\overline{\rm D}}}
\newcommand{\DroI}{{\overline{\rm D}}}
\newcommand{\eps}{\epsilon}
\newcommand{\vepsdi}{{\cal E}^\mathrm{d}_\mathrm{i}}
\newcommand{\vepsde}{{\cal E}^\mathrm{d}_\mathrm{e}}
\newcommand{\lraS}{\longmapsto}
\newcommand{\lra}{\longrightarrow}
\newcommand{\LRA}{\Longrightarrow}
\newcommand{\Equival}{\Longleftrightarrow}
\newcommand{\DRA}{\Downarrow}
\newcommand{\LLRA}{\Longleftrightarrow}
\newcommand{\diver}{\mbox{\,div}}
\newcommand{\grad}{\mbox{\,grad}}
\newcommand{\cd}{\!\cdot\!}
\newcommand{\Msun}{{\,{\cal M}_{\odot}}}
\newcommand{\Mstar}{{\,{\cal M}_{\star}}}
\newcommand{\Mdot}{{\,\dot{\cal M}}}
\newcommand{\ds}{ds}
\newcommand{\dt}{dt}
\newcommand{\dx}{dx}
\newcommand{\dr}{dr}
\newcommand{\dth}{d\theta}
\newcommand{\dphi}{d\phi}

\newcommand{\pt}{\frac{\partial}{\partial t}}
\newcommand{\pk}{\frac{\partial}{\partial x^k}}
\newcommand{\pj}{\frac{\partial}{\partial x^j}}
\newcommand{\pmu}{\frac{\partial}{\partial x^\mu}}
\newcommand{\pr}{\frac{\partial}{\partial r}}
\newcommand{\pth}{\frac{\partial}{\partial \theta}}
\newcommand{\pR}{\frac{\partial}{\partial R}}
\newcommand{\pZ}{\frac{\partial}{\partial Z}}
\newcommand{\pphi}{\frac{\partial}{\partial \phi}}

\newcommand{\vadve}{v^k-\frac{1}{\alpha}\beta^k}
\newcommand{\vadv}{v_{Adv}^k}
\newcommand{\dv}{\sqrt{-g}}
\newcommand{\fdv}{\frac{1}{\dv}}
\newcommand{\dvr}{{\tilde{\rho}}^2\sin\theta}
\newcommand{\dvt}{{\tilde{\rho}}\sin\theta}
\newcommand{\dvrss}{r^2\sin\theta}
\newcommand{\dvtss}{r\sin\theta}
\newcommand{\dd}{\sqrt{\gamma}}
\newcommand{\ddw}{\tilde{\rho}^2\sin\theta}
\newcommand{\mbh}{M_{BH}}
\newcommand{\dualf}{\!\!\!\!\left.\right.^\ast\!\! F}
\newcommand{\cdt}{\frac{1}{\dv}\pt}
\newcommand{\cdr}{\frac{1}{\dv}\pr}
\newcommand{\cdth}{\frac{1}{\dv}\pth}
\newcommand{\cdk}{\frac{1}{\dv}\pk}
\newcommand{\cdj}{\frac{1}{\dv}\pj}
\newcommand{\rad}{\;r\! a\! d\;}
\newcommand{\half}{\frac{1}{2}}
\newcommand{\ARZL}{\textquotedblleft}
\newcommand{\ARZR}{\textquotedblright}
\twocolumn[
  \begin{@twocolumnfalse}
  \title{The remnant of GW170817: a trapped neutron star with a hypermassive incompressible superfluid core}

\author{\thanks{E-mail:AHujeirat@uni-hd.de}~Hujeirat  A.A., \thanks{ravi.samtaney@kaust.edu.sa} Samtaney, R. \\
IWR, Universit\"at Heidelberg, 69120 Heidelberg, Germany \\
Applied Mathematics and Computational Science, CEMSE Division, KAUST, Saudi Arabia}
\maketitle
\begin{abstract}
Our bimetric spacetime model of glitching pulsars is applied to the remnant of GW170817.
Accordingly, pulsars are born with embryonic incompressible superconducting gluon-quark superfluid cores (SuSu-matter) that are
embedded in Minkowski spacetime, whereas the ambient compressible and dissipative media (CDM) are imbedded in  curved spacetime.
As pulsars cool down, the equilibrium between both spacetime is altered, thereby triggering the well-observed glitch phenomena.

 Based thereon and assuming all neutron stars (NSs) to be born with the same initial mass of $M_{NS}(t=0) \approx 1.25 \MSun,$ we argue that the remnant of GW170817 should be a relatively faint NS with a hypermassive central core made of SuSu-matter. The
effective mass and radius of the remnant are predicted to be 
$[2.8 \mathcal{M}_{\odot} < \mathcal{M}_{rem} \le 3.351 \mathcal{M}_{\odot} {]}$
 and  $R_{rem}=10.764$ km,
whereas the mass of the enclosed SuSu-core  is  $\mathcal{M}_{core}=1.7 \MSun.$ Here, about $1/2~ \mathcal{M}_{core}$
is an energy enhancement triggered by the phase transition of the gluon-quark-plasma from the microscopic into macroscopic scale. \\
  The current compactness of the remnant  is
$\alpha_c = 0.918,$  but predicted to increase as the CDM and cools down, rendering the remnant an invisible dark energy object,
and therefore to an excellent black hole candidate.  \\
\begin{center}
\end{center}
\end{abstract}
 \end{@twocolumnfalse}
 ]
 \section{Introduction}
 The astronomical event GW170817 marks the beginning of a new era of multimessenger astronomy, in which detectors of
gravitational waves and electromagnetic radiation operated almost simultaneously to follow up the merger event. Using LIGO detectors and advanced Virgo in combination with Fermi and INTEGRAL gamma-ray telescopes the event was localized to the galaxy 4993, in which two compact objects,
most likely two neutron stars, were found inspiraling and to subsequently merge and form a hypermassive compact object \cite[see][and the references therein]{AbbottNSProberties2019}. The component masses were inferred to lie between $[1.0\,\MSun \leq M_1, M_2 \leq 1.89\MSun]$ in the high spin case and $[1.16 \,\MSun \leq M_1, M_2 \leq 1.60\MSun]$ in the low case. Approximately $10\%$ of the total mass is predicted to have gone forming  an accretion disk, ejecta,  jets, X-ray flares and/or outflows \cite[see][and the references therein]{Shibata2017}.
\begin{figure}[t]
\centering {\hspace*{-0.35cm}
\includegraphics*[angle=-0, width=7.0cm]{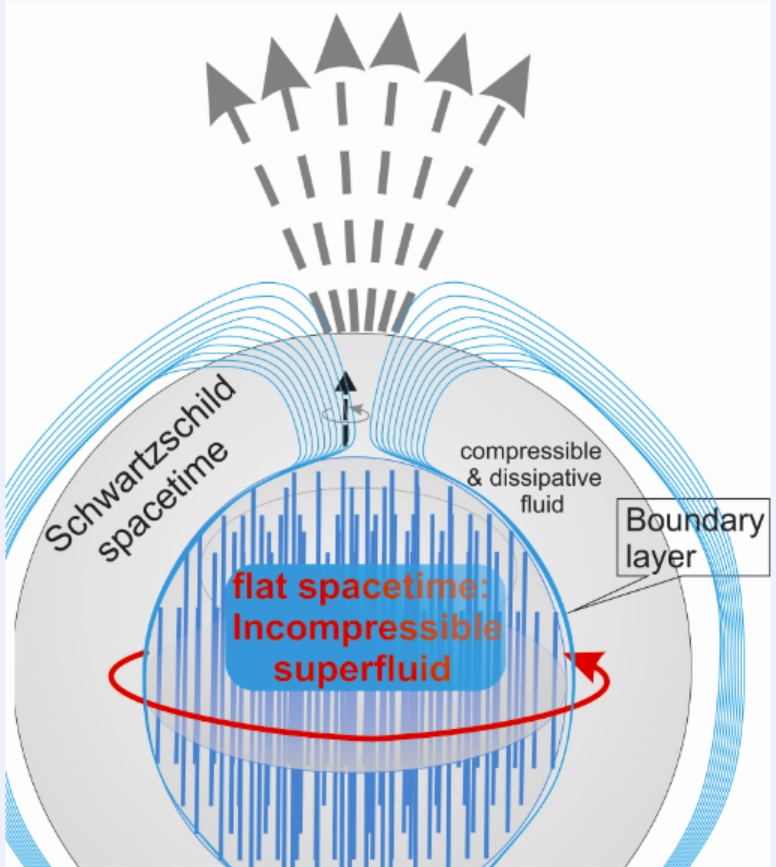}
}
\caption{\small A schematic description of the interior of the remnant of GW170817 as predicted by the bimetric spacetime scenario.
While the incompressible superfluid
in the core is enclosed in a Minkowski spacetime, the surrounding compressible and dissipative media
is set to be imbedded in a  curved spacetime. Both media are separated by geometrically thin  boundary layer, where the
pulsar's dynamo action is expected to operate.
}\label{NSInternal}
\end{figure}
While the exact nature of the remnant depends strongly on the final baryonic mass, it  may still
 fall in one of the following three categories:\\
\bit
\item[$\blacksquare$] \textbf{A hypermassive neutron star (HMNS)}\\
Based on observational and theoretical considerations, the possibility that the merger of the two NSs yielded  a short-living hypermassive
neutron star (HMNS) appears to be widely accepted \cite[see][and the references therein]{Margutti2018,PiroXray2019,Banagiri2020}. On the long-term however, it is not clear if such a massive object would survive a gravitational
collapse and end up as a  stellar BH \cite[see][and the references therein]{Gill2019}.  \\
On the other hand,  there are several observational signatures that obviously favor the formation a MNS. Namely,
the gamma-ray burst  GRB170817, which occurred 1.7 seconds after the merger event
 \cite{AbbottNSMergerLIGO2017}, the appearance of certain features after 155 days that may indicate  a reactivation of the central object. Moreover,
the formation  of a structured off-axis jet \cite{Margutti2018,Mooley2018},  the detected emission
powered by the radioactive decay of r-process nuclei synthesized in the ejecta \cite{Yu2018} and that a surface magnetic field
of order $10^{12}$ G is required to match the EM-radiation hint that the  remnant must be neutron-rich with  a hard
surface \cite[see][and the references therein]{PiroXray2019}.\\
\begin{figure}[t]
\centering {\hspace*{-0.35cm}
\includegraphics*[angle=-0, width=7.0cm]{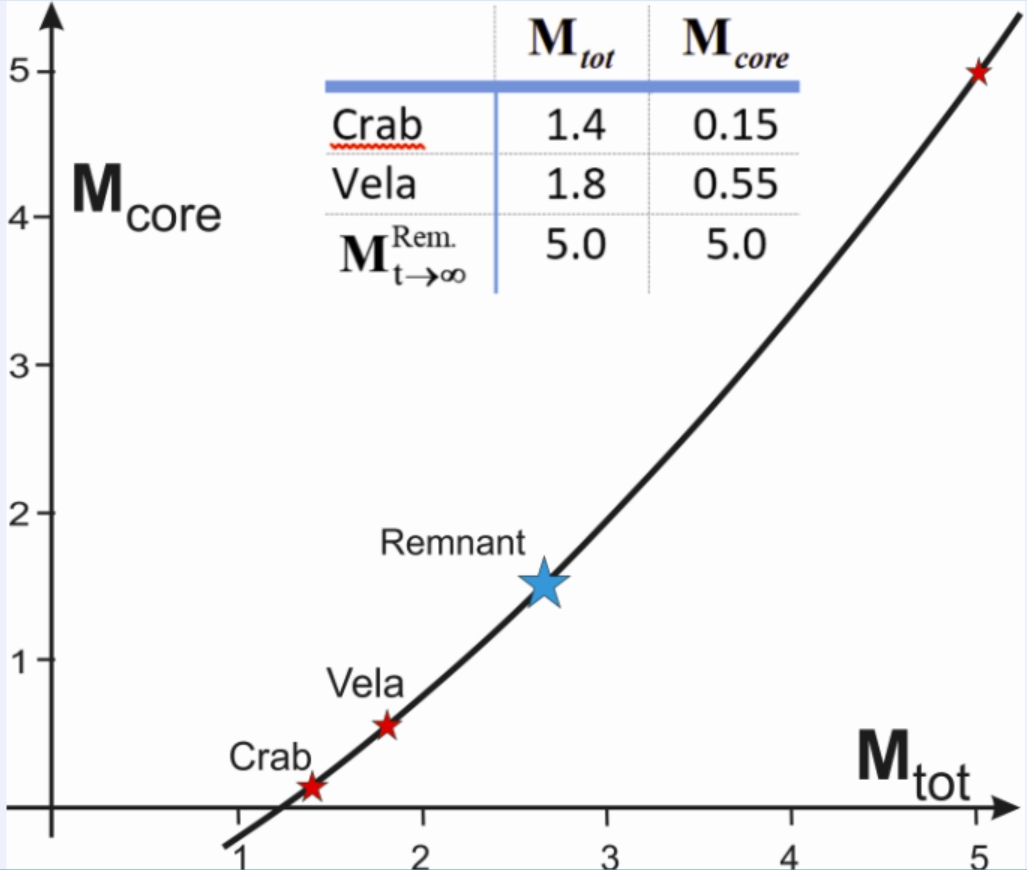}
}
\caption{\small  The total mass of the core (baryonic and dark energy) versus total mass of the entire object (: the mass of the core and
overlying baryonic shell) are displayed, using a polynomial interpolation of the values tabulated in the figure.  Part of these values
have been adopted from \cite{HujSam2020}.
}\label{Mass_Function}
\end{figure}
In particular, the observed steady brightening of GW170817 lasting less than 160 days after the merger, which is most likely powered by non-thermal synchrotron emission from plasmas propagating at  relativistic speeds, may be well-considered for ejecta from a central compact object with
hard surface. The Lorentz factors, $\Gamma s ,$ that correspond to plasma propagation in micro-quasar systems  generally fall in the range of
$[3 \leq \Gamma \leq 10 ] $, whereas $\Gamma s $ of the observed ejecta  in GW170817  hardly reach the lower range of this interval.\\

 \item[$\blacksquare$] \textbf{A stellar black hole (SBH)}\\
 Whether the remnant was a short or is a long-living HMNS,  its total mass is ${\mathcal{M}}_{tot} \geq 2.6 \MSun$ to $90\%$
confidence \cite[see][and the references therein]{AiMaxMass2020}.
In the absence of extraordinary stiff EOSs with exotic repulsive nuclear forces of unknown origin, the object should ultimately collapse into a  stellar black hole. Indeed, the revealed low  X-ray flux from GW170817 indicates weak magnetic field that is typical to stellar BHs \cite{Pooley2018}. Other arguments favoring the formation of a BH have been discussed, though the results obtained may be biased, as these studies assume a prior the
formation of a BH \cite[see][and the references therein]{Gill2019}.

However, it should be noted here that determining the nature of the remnant would require solving the time-dependent general relativistic Navier-Stokes
 equations with radiative transfer and magnetic fields at the background of a dynamically varying spacetime, which is beyond the state-of-the-art
simulations today. Also, the capability of receiving gravitational wave signals that carry information about the spin and compactness
of the NSs shortly before and of the remnant immediately after the merger event are outside the sensitivity range of LIGO.
\item[$\blacksquare$]   \textbf{Another type of a compact object }\\
In the absence of  direct and founded observational signatures that determine conclusively  the nature of the remnant, the possibility that
it might be neither a classical NS nor a stellar BH, but rather a new type of a compact object cannot be excluded.
The fact that the mass of the remnant falls in the range of $[2.5\, \MSun \leq \mathcal{M} \leq 5\, \MSun]$, where neither stellar BHs nor normal neutron stars have ever been observed, makes this conjecture  viable.   \\
In the following, we discuss this conjecture in detail and argue that the remnant is most likely made  of incompressible gluon-quark superfluid core embedded in a flat spacetime and surrounded by a shell of compressible and dissipative matter with a curved spacetime at the background as visualized
in Fig.(\ref{NSInternal}).

 \section{The remnant of GW170817 and its internal structure}
In the case of a quasi-static contraction of a massive NS: as  the event horizon, $R_\mathcal{H},$  and the star radius, $R_\star,$
approach each other,   the stiffness of the EOS and therefore the compactness must increase.  The sound speed should increase
and reach roughly  the speed of light at $R_\star= R_\mathcal{H} (1+ \epsilon), $  where $\epsilon$ is a sufficiently small number. Indeed, it was repeatedly
argued that the sound speed inside HMNSs must be larger than $c/\sqrt{3}$ \cite[see][and the references therein]{BedaqueSoundVel2015}.
A possible limiting EOS may have the form:
\beq
     P_L = n^2 \DD{\D}{\D n} ( \large{\varepsilon}/n)
 \xrightarrow[]{\large{\varepsilon} \rightarrow a_0 n^2}
  \large{\varepsilon},
\eeq
 where $a_0,~n,~ P_L, ~\varepsilon$ denote a constant coefficient,  number density, local pressure and the density of internal energy, respectively.  Here the rate of interaction of
sub-nuclear particles, namely mesons and gluons,  reaches the saturation limit, at which the chemical potential, i.e. energy per particle, attains
its  universal maximum value, which was predicted to be around $ n_{cr} \approx 3 \times n_0,$ where $n_0$ is the  nuclear number density \cite{HujeiratMassiveNSs18}. In this case,
\beq
\begin{split}
 \large{\varepsilon}= a_0 n^2 \xrightarrow[n \rightarrow n_{cr}]{} a_0 n^2_{cr}~ & = \large{\varepsilon}_{cr} \\
                                                                                                                                  &   = \mathcal{O}(10^{36})\,erg/cc
\end {split}
\label{EdConstValue}
\eeq
In the present study, this EOS governs fluids that are the maximally compressible or  purely incompressible.
Under these conditions, the constituents are capable of resisting all kinds of external perturbations, including further contractions by gravity.
However an  EOS of the type $\large{\varepsilon}= P_L = const.$ is incompatible with traceless
mass-energy tensors (METs)  and therefore does not obey conformal invariance \cite{BedaqueSoundVel2015}.
Similar to weakly compressible terrestrial fluids,
 the pressure in incompressible fluids loses its local thermodynamical character and turns into a mathematical term only\footnote{in terrestrial weakly incompressible fluid the pressure is treated as  a Lagrangian multiplier.},
 which nevertheless must be chosen, so to ensure that $dP_L/d\large{\varepsilon} \leq 1.$\\

In fact, fluids governed by the limiting EOS: $~~P = \large{\varepsilon}~(> \large{\varepsilon}_0)$ cannot accept stratification by gravity or, equivalently by the
curvature of spacetime. Indeed, it was recently verified that  once the matter at the very center of massive NSs become purely incompressible,
then the embedding spacetime must flatten and become conformally flat \cite[see][and the references therein]{HujeiratGW170817}.  The total energy of a stationary and zero-stratified supradense nuclear matter should consist of the rest energy only, as otherwise, thermal, magnetic and kinetic energies can be affected
directly by gravity. Thus  the energy state of purely incompressible supradense nuclear fluids should correspond to the universal lowest-possible
energy state.
 Based thereon, our argument may be summarized as follows:
\ben
  \item Under certain conditions, supradense nuclear fluids  may become maximally compressible, i.e. purely incompressible,
          at  which the energy density attains  the universal  maximum value  \textbf{$\large{\varepsilon}_{max}.$ }Here the interaction rate between the
           constituents saturates  as they communicate with each other at the  speed of light. Energy divergence is prohibited
           as massless gluons are the mediators whilst the quarks must motionless in space,  but oscillatory in time.
          These conditions should apply to the gluon-quark-plasma inside individual baryons at zero-entropy.
\item  The classical pressure in purely incompressible supradense nuclear fluids loses its local thermodynamical character.
          In fact, as the constituents communicate with each other at the speed of light, the interaction-power is sufficiently strong to smooth
            out all possible potential barriers between individual baryons, thereby enhancing their merger and forming
          an ocean of gluon-quark superfluid.
\item  Purely incompressible supradense nuclear matter governed by the EOS,
        \beq
              P = \large{\varepsilon}_{max}= const.
        \eeq
          should have zero-entropy and embedded in a flat spacetime. This strong conjecture raises the possibility that
         entropy of matter and curvature of spacetime  may have  hidden connections.
\een
\eit
 \section{Basic assumptions and the solution strategy }
Very recently, we applied the  bimetric spacetime scenario of glitching pulsars to investigate the internal structures  both of
the Crab and Vela pulsars \cite{HujSam2019}. The model was capable of re-producing and explaining several mysterious features that observed to accompany
both pulsars, namely the deriving mechanisms underlying the glitch phenomena  and their rate of reoccurrence, the origin of under/overshootings observed to associate their glitch events
as well as their cosmological fate.
Based thereon,   newly born pulsars are predicated to have the initial mass of $ 1.25 ~\MSun$ and  an embryonic SuSu-core  of $ 0.029~ \MSun$
to evolve into a Crab-like pulsar after 1000 years and subsequently into a Vela-like pulsar 10,000 years later to finally fade away as an invisible dark energy object  after roughly 10 Myr. The cores of both pulsars were predicted to have the masses: $\mathcal{M}^{Crab}_{core} = 0.15 \MSun$ and $\mathcal{M}^{Vela}_{core} = 0.55 \MSun$  \cite{HujSam2020}.\\

In order to apply the  model to the remnant of GW170817, the following assumptions must be made:
\begin{itemize}
  \item[-]  The masses of the two merging NSs  read: $\mathcal{M}^{1}_{tot} =\mathcal{M}^{2}_{tot} = 1.4 \, \MSun,$  which fall  in the mass-range revealed by observations \cite{AbbottNSProberties2019}. Although our results  are not too sensitive to these
      exact values, both masses are remarkably close to that of the Crab pulsar and therefore we may assume that both NSs should have the same
 cosmic evolution.
      Specifically, both NSs should  have identical initial conditions,  are perfectly isolated with no energy loss or gain from or to
    their surroundings. As in the case of the Crab and Vela pulsars, any mass-difference is due to age-difference: as the object ages,
     it becomes colder, the inertia of its SuSu-core increases, which is associated with a topological change of spacetime embedding
    its interior. \\
The total mass of the object consists of core's mass, half of which is baryonic matter while the other
half is due to the  enhanced gluon cloud enclosing the quarks on the macroscopic scale.  The other contribution to  $\mathcal{M}^{1}_{tot}$ comes from the ambient normal fluid.\\
Based thereon each of two the NSs prior merger should have  approximately the same age, same total masses
     as the Crab pulsar and therefore the same core's masses: $\mathcal{M}^{Crab}_{core} = 0.15 \MSun$ \cite[see][for further details]{HujSam2020}.

 \item[-] The incompressible superfluid cores of both objects should have survived the violent merger and preserved their inertia.
  While this assumption appears to be too strong, it may be supported by the following arguments:
             \begin{itemize}
               \item Similar to gluon-quark plasma inside individual baryons, the incompressible gluon-quark superfluid inside the cores of
                pulsars cannot live in free  space, but hidden behind a confining quantum barrier. The effective energy of the potential barrier
              is sufficiently strong  to protect the core against deformations by tidal forces. Indeed,  the sound crossing time of each core
              is predicated to be one million times shorter than the dynamical time scale during the inspiraling phase of both NSs prior merger.
               \item In QCD, the bag energy confining  gluon-quark plasmas inside individual baryons is of order 300 MeV. When N
               baryons  are set to merge together under zero-temperature and pure incompressibility conditions, then the  size, V,  and mass
              of the resulting super-baryon must increase  with N, i.e. $dV/dN = const.$ Recalling that roughly $99\%$ of the effective energy of an individual baryon is due to the gluon field, then the energy of the gluon cloud confining the quarks
              inside the super-baryon must  moderately increase with N, i.e. an energy enhancement  of the type:
                $d\varepsilon^{gluon}/dN = \alpha_0 \, N^\beta$ is forbidden as  $N$ is typically of order  $ \mathcal{O}(10^{57}),$ so that
                 any value of $\beta >0$  would lead energy divergence.
                 A reasonable prediction would be that the merging process of N-neutrons at zero-temperature
                is energetically upper-bounded by the  constant: $d\varepsilon^{gluon}/dN = 2 \varepsilon_0,$ where $\varepsilon_0$ is the
                rest energy of a neutron.

                One may conjecture that the extra energy originates from the gluon field  necessary for  stably confining the quark-ocean
               inside the super-baryons and shield it from the outside universe \cite{Witten1984}.  Indeed, the effective energy stored in the
               creation of the short-living pentaquark that was detected in the LHC experiment  was found to be about 4.5 GeV, which is
                 consistent with our scenario \cite{LHCb2015}, though the physical conditions are completely different: while the plasma at the LHC
                  is characterized by very high temperature  and extremely low density, the matter in the central cores of pulsars
                  should be supradense nuclear  with temperature  much lower than the corresponding Fermi one.
               \item The incompressible supranuclear dense matter inside the cores of NSs must have the lowest possible energy state and behaves as a single quantum entity. Such a fluid is expected to be well-equipped to resist all types of external perturbations, including tidal deformation. In this case,  the only source  left for energy emission  would be the ambient compressible and dissipative nuclear fluid in the shell.
              \end{itemize}
\end{itemize}
\begin{figure}[t]
\centering {\hspace*{-0.35cm}
\includegraphics*[angle=-0, width=7.0cm]{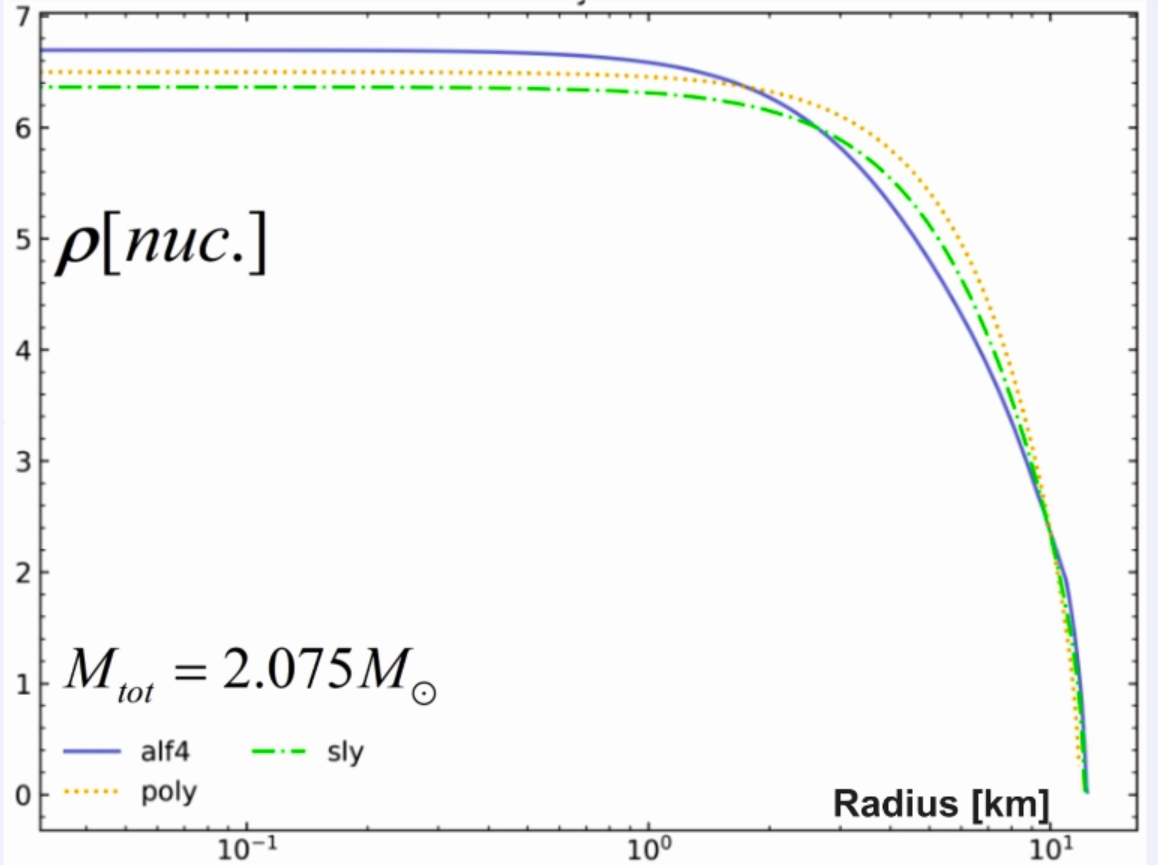}
}
\caption{\small The internal structure of a marginally stable massive NS using the three different EOSs:  alf4(\cite{Aford2005}),
sly (\cite{Douchin2001}) and poly (:the polytropic EOS in which the coefficient $\mathcal{K}$ and
 $\gamma$ were optimized to fit the other two profiles). Except poly, non of the EOSs, including other 12
 models, where capable of stably modelling the interiors of NSs  more massive than $2.075 \MSun.$
}\label{EOS3}
\end{figure}

\begin{figure}[t]
\centering {\hspace*{-0.35cm}
\includegraphics*[angle=-0, width=7.0cm]{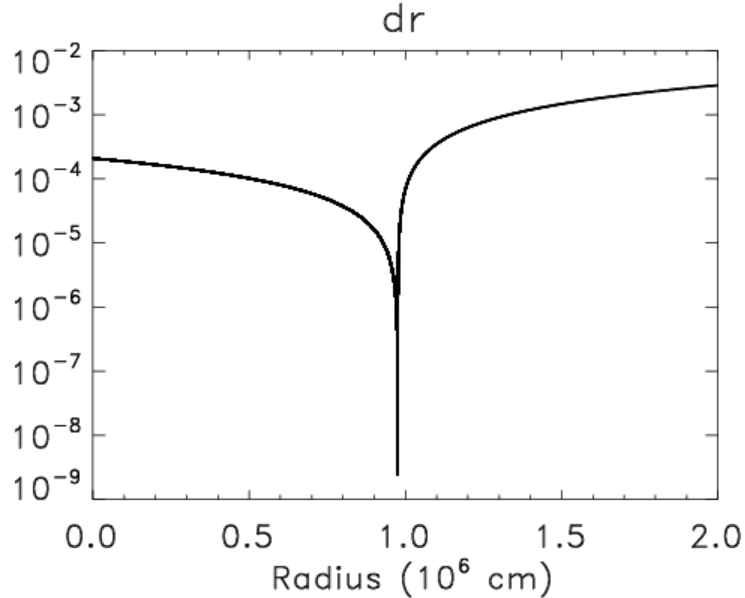}
}
\caption{\small The non-linear distribution of the grid spacing  versus radii. Here the density of
grid points was optimized in order to capture the fine structures of the remnant in the vicinity of its surface.
}\label{dxDistribut}
\end{figure}
 \section{The numerical approach}
Based on the above-mentioned arguments, the mass and size of the remnant's core  may be obtained here by extrapolating the sizes of the cores of the
well-studied  Crab and of the Vela pulsars as well as recalling that the remnant should turn invisible, once the object has metamorphosed entirely into a SuSu-object. Thus a NS with the initial mass of $M_0 = M(t=0)=2.5~\MSun$ will double its mass at the end of its visible lifetime, i.e.
$M(t=\infty) = 5~\MSun.$\\
In Fig. (\ref{Mass_Function}) we show the distribution of effective masses of the cores versus the total masses of the objects. Here we use the tabulated values shown
 in Fig. (\ref{Mass_Function}) to construct the quadratic mass-function in units of $\MSun$:
\beq
\begin{split}
   \mathcal{M}_{core} = 0.108507\,\mathcal{M}^2_{tot} &+ 0.652778  \, \mathcal{M}_{tot}  \\
                                                                                         &  - 0.976562,
\label{Massfunction}
\end {split}
\eeq
which applies for $ \mathcal{M}_{tot} \geq 1.25\MSun.$\\
Let $t=t_0$ denote the time, at which the two NSs merged together. Based on observations the total mass of both objects is: $M_{tot}(t=t_0-\epsilon_{t_0}) = 2.8\,\MSun,$ which
consists of $ 2.5\,\MSun$ baryonic matter and $ 0.3\,\MSun$ of dark energy, where
$t=t_0-\epsilon_{t_0}$ corresponds to the time shortly before the merger event and  $\epsilon_{t_0}$ denotes a short instant.\\
Based on the mass-function $ \mathcal{M}_{core},$  the core's  mass of the remnant appears to have undergone a dramatic increase
during the merger  from   $0.3  \MSun $ to $1.7 \MSun,$ which implies an additional input of dark energy that  amounts to $0.55 \MSun,$
yielding  a massive NS of $3.351 \MSun,$ and consisting of $2.5 \MSun$ in the form of baryonic matter and $0.85 \MSun$  of dark energy.\\
The mechanism underlying the injection of dark energy into the system here may be explained as follows:\\
When the matter at the center of a massive NS is compressed by the surrounding curved spacetime, the separation between two arbitrary
 baryons, $\ell_b,$ may decrease down and become comparable to the mean free path $\ell_g,$
between quarks in individual baryons and at zero-temperature. In this case a free energy $\triangle \varepsilon^+$ of order $c\hbar/2\ell_g$
would be generated and may easily surpass the $0.939$ GeV limit, beyond which  baryon's merger becomes a viable process. $\triangle \varepsilon^+$ is expected to be stored locally, and specifically  to enhance the surface tension needed for confining the enclosed
ocean of quarks. At certain time, this energy would be  liberated,  once the core has decayed into a pure baryonic matter.

\begin{figure}[t]
\centering {\hspace*{-0.35cm}
\includegraphics*[angle=-0, width=7.0cm]{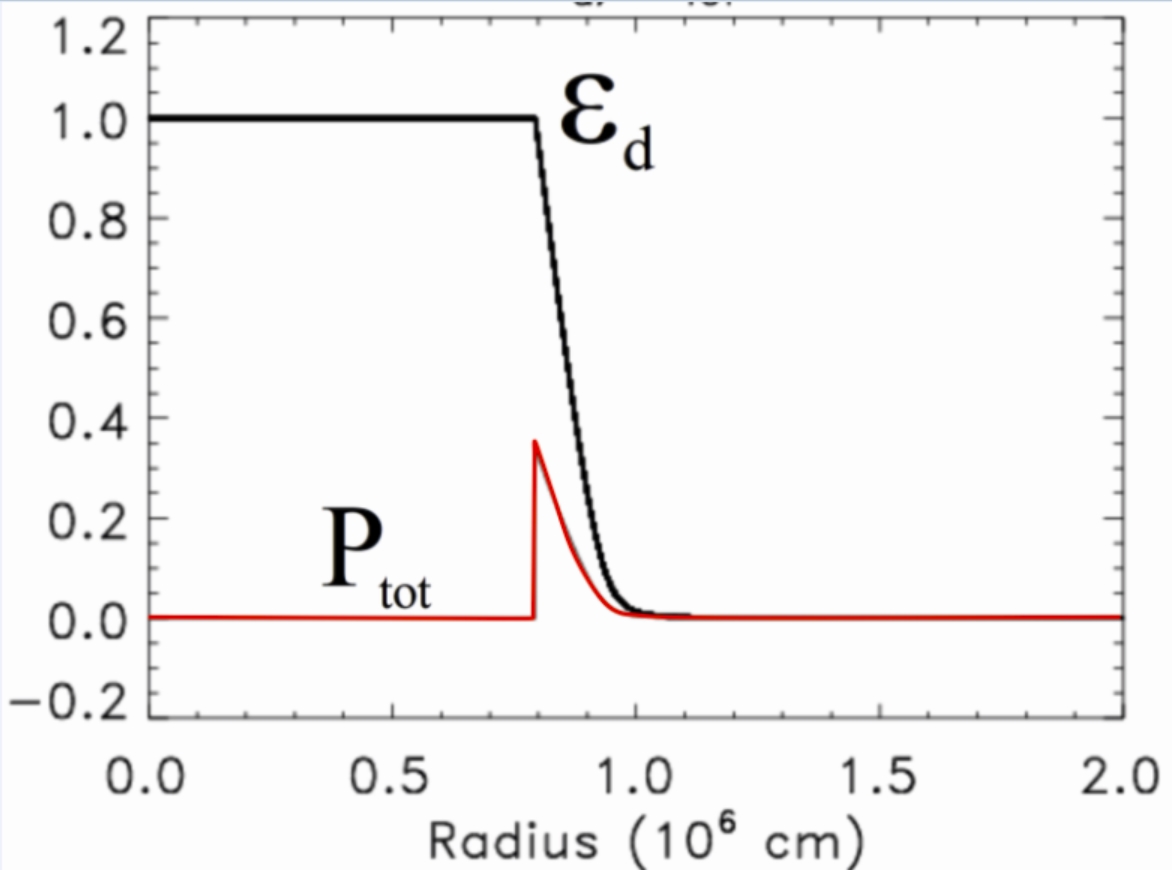}
}
\caption{\small The profiles of the energy density $\varepsilon_d $ and  the total pressure, $P_{tot},$ versus radius in units of  $2~ \varepsilon_{cr}$.
Inside the core, where  the spacetime is flat, the energy density attains the constant value $\varepsilon_d = 2~ \varepsilon_{cr}$
 ( see Eqs. \ref{EdConstValue}, \ref{PVacuum1} and  \ref{PVacuum2}).
Outside the core, where the matter is compressible and
dissipative, the radial-distribution of $\varepsilon_d$ is determined by solving the TOV equation at the background of a
Schwarzschild spacetime using the "poly"  EOS.    The total pressure inside the core is the superposition of the normal local pressure
 and the negative pressure due to vacuum.  While $P_{tot}$ vanishes inside the core, it becomes equal to the local pressure
outside it.
}\label{EdPtot}
\end{figure}
\begin{figure}[t]
\centering {\hspace*{-0.35cm}
\includegraphics*[angle=-0, width=7.0cm]{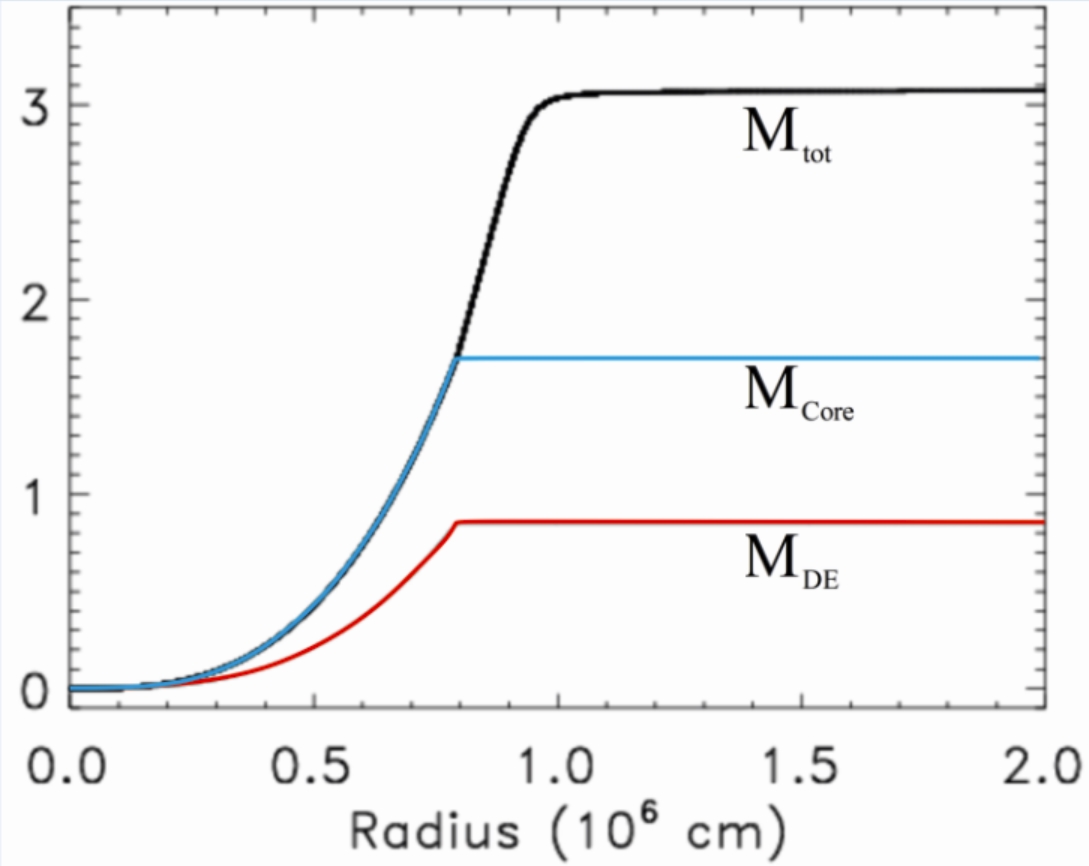}
}
\caption{\small
The radial-distributions of the total effective mass of the remnant  $M_{tot},$
 the  total mass of the core, $M_{core},$  and the  contribution of dark energy,  $M_{DE},$ to the effective mass of the core.
The final masses here read:  $M_{tot}= 3.351\MSun,~M_{core} =  1.7\,\MSun$ and $M_{DE} =  0.85\,\MSun.$
}\label{Masses}
\end{figure}
Based thereon, the Tolman-Oppenheimer-Volkov equation (TOV) was solved, using  the core mass and its energy-density as input parameters.
In the present case, the inertia of the core is expected to significantly enhance the curvature of
spacetime and therefore the distribution of matter in the shell surrounding  the core. The matter here is set to obey the polytropic EOS:
 $P= \mathcal{K}\rho^\gamma.$  In order to obtain the optimal values of  $\mathcal{K}$ and $\gamma,$
the TOV equation was solved for the maximum possible mass that can be obtained using the EOSs  "alf4" and "sly" \cite{Douchin2001,Aford2005}.
As shown in Fig. (\ref{EOS3}), the resulting mass turns to be: $\mathcal{M}^{max}_{NS} = 2.075\,\MSun, $ which is approximately equal  to that
of  PSR J0348+0435.\\
Interestingly enough, using "alf4" and "sly"  EOSs, the central density obtained here roughly equal to the universal maximum value:
$\rho_{max} = 6\times\rho_0,$  at which the fluid turns purely incompressible superfluid. \\
The fitting procedure of the polytropic EOS with  alf4" and "sly"  ( see Fig. \ref{EOS3}), yielded the following values:
\beq
\barr{lll}
\mathcal{K} & =& 1.98183\times10^{-6}  \\
 \gamma  &=& 2.72135.
\label{PolyPar}
\earr
\eeq
Assuming $\rho_{max} $ to exist in our universe, then the physical conditions governing both the cores of MNS  and
  of the remnant of GW170817, must be identical.\\

\begin{figure}[t]
\centering {\hspace*{-0.35cm}
\includegraphics*[angle=-0, width=7.0cm]{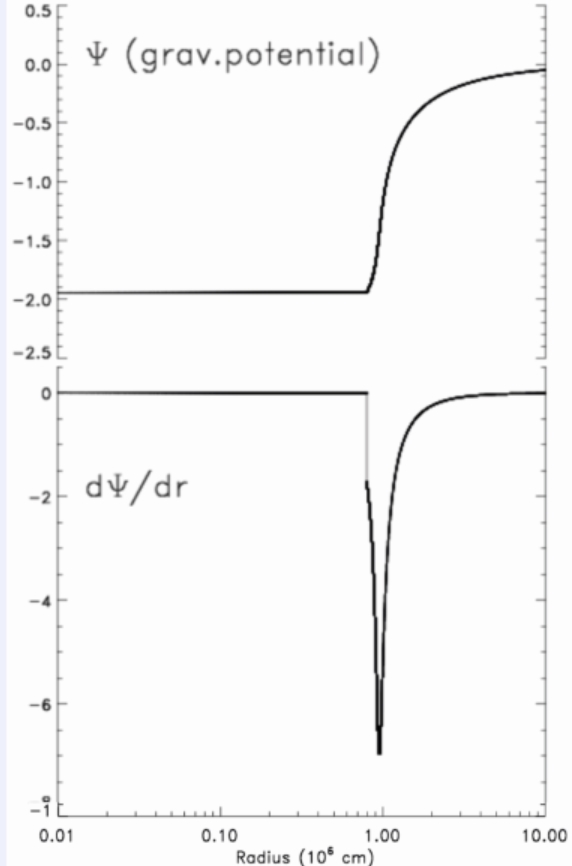}
}
\caption{\small The radial distribution of the gravitational potential, $\Psi,$ and its radial derivative throughout the remnant and
the surrounding space.
}\label{gravPot}
\end{figure}

\begin{figure}[t]
\centering {\hspace*{-0.35cm}
\includegraphics*[angle=-0, width=7.0cm]{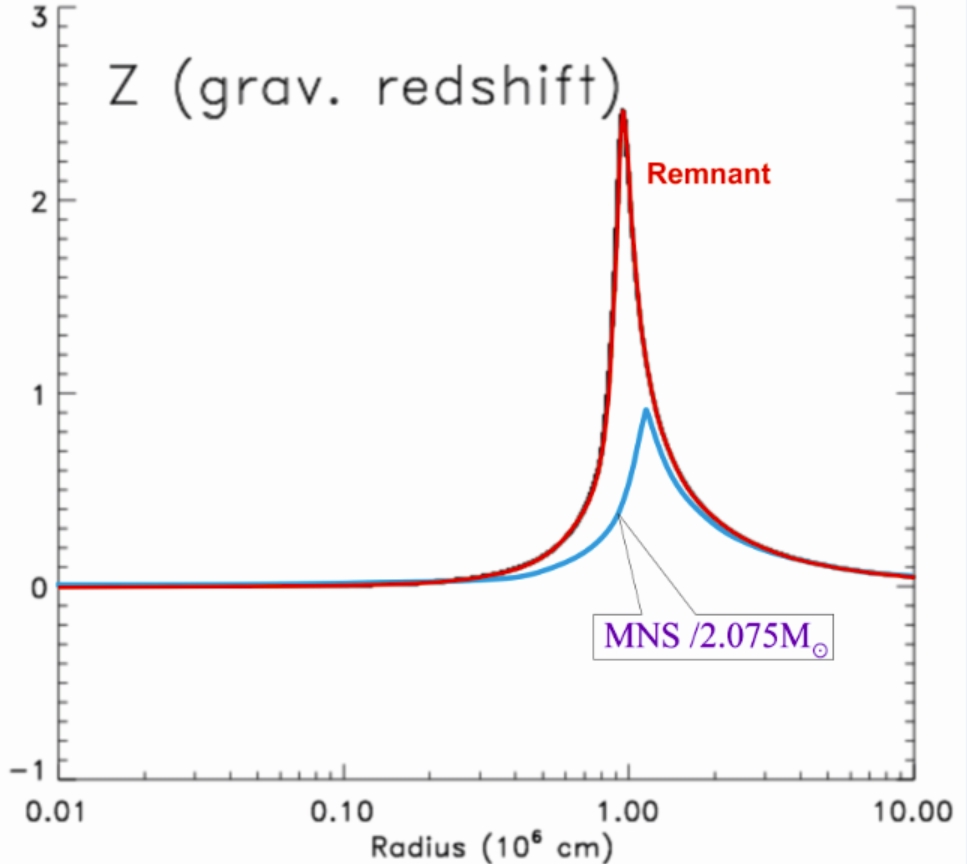}
}
\caption{\small The radial distribution of the gravitational redshift, Z, throughout the remnant and
the surrounding space (red-colored) to be compared to that of a massive NS (MNS/blue-colored) with $M_{NS} = 2.075~\MSun.$ Z at the
surface of the remnant is roughly four times larger than Z at the surface of the MNS.
}\label{GravRedShift}
\end{figure}

In order to accurately capture the strong gradients in the vicinity of the remnant's surface, the explicit adaptive mesh
refinement (EAMR) method  has been employed \cite{HujSam2019}.  Here approximately $10^5$ grid points have been used reaching
an aspect ratio of $dr_{max}/dr_{min} \approx 10^5$  (see Fig. \ref{dxDistribut}).
In Fig.(\ref{EdPtot}) the profiles of the normalized total energy density and the total pressure throughout the entire object versus radius are shown. These are defined as follows:
\beq
\begin{split}
     \varepsilon_d =  \varepsilon_{bar} +  \varepsilon_{\phi}, & ~~~~~~~~~P_{tot} =  P_{bar} +  P_{\varphi},\\
\end{split}
\label{PVacuum1}
\eeq
where
\beq
\begin{split}
   \varepsilon_{\phi} = \DD{1}{2}\dot{ \phi}^2 + V(\phi)+ \DD{1}{2}( \nabla \phi)^2  & \\
  P_{\phi} = \DD{1}{2}\dot{ \phi}^2 - V(\phi) - \DD{1}{6}( \nabla \phi)^2.
\end{split}
\label{PVacuum2}
\eeq
Here  $\phi, V(\phi)$ generally denote the scalar field and its interaction potential  with the baryonic matter. In the present case,
 $\phi$ is a spatially averaged gluon field and  $V(\phi)$ is  the potential energy of the gluon-cloud embedding
the quarks inside the super-baryon. The subscript "bar" denotes the contribution due to baryonic matter whereas $\varphi$ due to vacuum.
The scalar field,  $\phi,$ is assumed to be significantly enhanced when undergoing a
phase transition from the micro into the macroscale. Such enhancement is necessary in order to maintain the super-baryon stable,
while hiding it from the outside world \cite[see][and the references therein]{Hujeiratmetamorph2018}.
Inside the core, the incompressible superfluid is stationary and therefore $\phi$ is spatially and temporary constant, i.e.
\beq
                          \dot{\phi} = \nabla  \phi = 0.
\eeq
In this case $V(\phi)=\varepsilon_{bar},$ which is equal to the effective resistive pressure, $P_L,$ required to maintain
the quarks apart at a minimum distance $\ell_{min}$ which is of order $0.6~$ fm.\\
Consequently, inside the core we have $\varepsilon_d = 2\, \varepsilon_{bar},$ but a vanishing total pressure $P_{tot}.$
Outside the core, however,   $\varepsilon_d$  runs as dictated by the TOV-equation at the background of a Schwarzschild spacetime and
using the polytropic EOS with the parameters specified  in Eq.(\ref{PolyPar}).\\
Following Eq.(\ref{Massfunction}), the total mass of the core is $1.7\,\MSun$, half of which is made of baryonic matter and the other  half is due to
the macroscopic enhancement of the gluon field (see Fig. \ref{Masses}). Using the polytropic EOS with the parameters given in Eq. (\ref{PolyPar}),
the TOV equation was integrated to determine the radius and therefore the compactness of the remnant. Our high-resolution numerical
calculations yielded the  radius  of $R_{rm}=10.764$ km for a remnant  mass of $3.351~\MSun$ to give rise to  a compactness parameter:
$\alpha_c = 0.918$  as shown in Figs. (\ref{Masses}) to  (\ref{GravRedShift}).\\
Obviously, the predicated value of the remnant's radius here is consistent with the recent investigations \cite[see][and the references therein]{Capano2019}.

In Fig. (\ref{gravPot}) we show the Schwarzschild matric exponent, $\Psi,$ which is, in the limit of weak gravitational field, reduces to the gravitational
potential.  Inside the core, where the embedding spacetime  is flat,  $\Psi$  attains a negative constant value, but increases abruptly throughout
the ambient medium and goes to zero at infinity. Across the surface, $\Psi$  undergoes a dramatic spatial variation. This can be seen from $d\Psi/dr, $ which vanishes
inside the core, but then decreases and increases dramatically in the vicinity of the remnant`s surface. Thus free particles in this region would
experience  rapid acceleration and deceleration, thereby giving rise to radiation emission  that are strongly redshifted and may appear
four times fainter than those emitted from the surface of massive and highly compact  NSs  (see Fig. \ref{GravRedShift}).

 \section{Summary}
 In this paper we have presented a model for the remnant of GW170817, which is based on the bimetric spacetime scenario of
glitching pulsars.  Accordingly, the core is made of incompressible superconducting,  gluon-quark superfluid (SuSu-matter) embedded in
a flat spacetime, whereas the ambient compressible and dissipative medium is set to be imbedded in a Schwarzschild spacetime.\\

The present study relies on our previous investigation of the cosmic evolution of the Crab and Vela pulsars as well as on
the topological change of spacetime inside pulsars as the driving mechanism  for the glitch phenomena in pulsars.\\
Based thereon we have found that the remnant must have a total mass of  $\mathcal{M}^{rem}_{tot} = 3.351\, \MSun,$ a radius $ R_{rem} = 10,764$
km and it should enclose a hypermassive SuSu-core of $\mathcal{M}^{rem}_{core} = 1.7\, \MSun.$ The matter inside the shell surrounding the core
is compressible and dissipative and its stratification is dictated by the curvature of the embedding Schwarzschild spacetime.  The massive object is deeply trapped in spacetime and the radiation emitted from its surface would be redschifted, reaching   $Z \approx 3,$ i.e. approximately 5 to 7  times more
fainter than "Z"s at the surfaces of  normal NSs.\\
Thus the remnant is actually a NS, which has a hypermassive SuSu-core overlaid by a neutron-rich matter.
On the cosmological time scale, when all secondary energies in the shell has exhausted\footnote{Energies other than the rest one.}, its compactness would saturate around one,
thereby turning the object invisible and making it an excellent black hole candidate.\\
The dark energy in the SuSu-core of the remnant is predicated to be  $0.85\, \MSun.$ This energy enhancement
originates from the phase transition of the gluon field from microscopic scale inside separated baryons into the macroscopic scale, where the core behaves as a single quantum entity,but still hidden from the outside universe.  \\
However, while the model is based on  several reasonable assumptions, work must be still done to verify their validities,  specifically:
\begin{itemize}
  \item  The existence of a universal maximum density and the state of pure incompressibility
  \item  The origin of gravitation and its hidden connection to entropy
  \item  The viability of the bimetric spacetime scenario inside pulsars
  \item  The mechanisms underlying the glitch phenomena in pulsars and young neutron stars.
\end{itemize}

Finally, similar to newly born pulsars, a significant part of the rotational energy prior to the merger should have been stored both
in the core and in the overlying shell of the remnant. The resulting configuration is expected to considerable deviate from
spherical symmetry and therefore serves a rich source for the emission of gravitational radiation with $\nu > 10^3$ Hz.
However, this range of frequencies is far beyond the current sensitivity range of LIGO.\\

Moreover, when correlating the total entropy production to the revealed radiation power from GW170817, we find that the  entropy is much
below the expected value from an object of $2.78\,\MSun,$ irrespective whether the remnant is a massive NS or a
stellar BH \cite{PiroXray2019}. Indeed,  a  massive entropy-free superfluid core  may nicely explain the origin of the entropy deficiency here.

Assuming the next generation of high resolution observations would render our scenario reasonable, then current scenarios related to the
fate of the first generation of NSs, the relatively small number of NSs and BHs in the Galaxy, the origin
of dark matter and dark energy in the universe must be revisited.

 \end{document}